\documentclass[a4paper,10pt]{amsart}

\usepackage{amssymb,epsfig,amsmath,amsthm}
\numberwithin{equation}{section}
\begin{document}

\title[The Blockchain]{{\Large The Blockchain: A Gentle\\ Four Page Introduction}}

\date{November 2016\vspace{0.5cm}\\
Record Currency Management, Windsor, UK}

\author{J. H. Witte}
\address{Record Currency Management, Windsor, UK}
\email{jwitte@recordcm.com}

\begin{abstract}
Blockchain is a distributed database that keeps a chronologically-growing list (\emph{chain}) of records (\emph{blocks}) secure from tampering and revision. While computerisation has changed the nature of a ledger from clay tables in the old days to digital records in modern days, blockchain technology is the first true innovation in record keeping that could potentially revolutionise the basic principles of information keeping. In this note, we provide a brief self-contained introduction to how the blockchain works. \\

\end{abstract}

\maketitle

\vspace{-0.75cm}
\section{Introduction}

A \emph{record} is any piece of evidence about the past -- especially an account kept in writing or some other resilient form -- which facilitates the accurate preservation of information through time. Traditionally, a specific and regularly updated record in paper or book form is referred to as a \emph{ledger}.

For any two individuals engaging in an information-dependent correspondence or exchange (as is almost always the case in business), record keeping presents a risk. Unless\vspace{-0.3cm}
\begin{itemize}
\item[a)] both parties are happy to rely on the records of a central authorised entity,
\item[b)] or one of the parties is happy to rely fully on the record keeping of the other party (which is rather rare in business),\vspace{-0.3cm}
\end{itemize}
careful comparison and matching of two independently held sets of records is the only possible solution.

Blockchain technology promises a distributed database that keeps a continuously-growing list (\emph{chain}) of records (\emph{blocks}) secured from tampering and revision, making comparison and matching of any separately held records unnecessary. If successful, blockchain could therefore substantially facilitate (and therefore accelerate) almost every multi-party decision-making process, including business, governmental, and private.

For example, the avoidance of past tampering of a transaction trail ensures that no new money has been unlawfully created. For cash, this is done through the use of bank notes that cannot (or are very hard to) be forged, and for electronic accounts, this is done by only allowing trusted participants (normally banks) to store balances. Evidently, it is central authorised entities that keep a ledger and prevent our financial systems from falling into chaos. Blockchain technology proposes to use a distributed (and therefore democratic) confirmation process of financial transactions instead.

\section{Technical Prerequisites}

A blockchain is designed based on two well-established mathematical ideas.

 \subsection{Public Key Encryption}
 
 Modern cryptography relies on the use of a \emph{public} and a \emph{private} key. Alice can share her public key freely with anyone (and maybe even publish it on her website), while her private key is known only to her. If Bob encrypts a message using Alice's public key, then no one except for Alice with her private key can ever decrypt the message -- as long as Alice keeps her private key safe, the communication between the two is unbreakable.
  
 \subsection{Hash Keys}

A \emph{hash key} is any fully-defined function which takes an alpha-numeric sequence (i.e., a string of letters and numbers) of arbitrary length and reduces it to one of predefined finite length. Even though, in theory, this means that duplication (i.e. more than one input resulting in the same output) is possible, for modern hash keys this can be assumed to be extremely unlikely.

One notable feature of hash keys is that they are (almost) impossible to invert. For example, if Bob is an internet provider who stores only the hash keys of user passwords (rather than the passwords themselves), than he can check Alice's user-login correctly without even knowing Alice's actual password. 

\section{Bitcoin: The Beginnings}

In his 2008 paper \emph{Bitcoin: A Peer-to-Peer Electronic Cash System}, a person calling themselves Satoshi Nakamoto (the name is a pseudonym and the real author is unknown) laid the foundation for the algorithmic network behind the crypto-currency \emph{bitcoin}. Even though the bitcoin itself continues to function and circulate successfully, Nakamoto's ideas behind the creation of the bitcoin have since exceeded the original application by far and are now known independently as the \emph{blockchain}.

Nakamoto created the bitcoin by inventing the following architecture.
\begin{itemize}
\item[i.] We have a distributed network of participating agents all of whom store a copy of the blockchain. The blockchain (the \emph{distributed ledger}) is a record of all \emph{transactions} to date. (This is very important: the blockchain stores transactions, not balances.)\vspace{0.25cm}
\item[ii.] Participating agents also hold bitcoin wallets, which represent ownership of any received bitcoin payments. More specifically, every bitcoin transaction is represented in the blockchain as the public key of a cryptographic public/private key pair, while the transaction's private key is held in someones bitcoin wallet. Anyone in possession of the blockchain can use a transaction's public key to encode messages which only the transaction's rightful owner can decrypt correctly.

For example, if Alice tells Bob that she owns 30 bitcoins (which she earlier received in a transaction from Charlie), then Bob (or anyone else) can instantly confirm Alice's claim to ownership by asking encrypted questions -- unless Alice truly owns the private key she claims to have, she cannot reply correctly.
\end{itemize}
Up to this point, the described design creates a \emph{static} distributed ledger. But imagine that now Alice wants to buy Bob's bike for 30 bitcoins, but that Bob is only willing to hand over the bike once both of them can independently confirm Alice's payment to Bob from their copies of the blockchain. To make the transaction possible, we have to extend and update the blockchain.

\begin{itemize}
\item[iii.] To initiate the transfer of 30 bitcoins to Bob, Alice requests a new public key from Bob and then sends her proposed transaction to all agents in the network.\vspace{0.25cm}
\item[iv.] For Alice's proposed transaction to be approved, one agent in the network now needs to
\begin{itemize}\vspace{0.25cm}
\item[a)] confirm that Alice is the rightful owner of the 30 bitcoins she intends to transfer (see ii. above),
\item[b)] and create a new hash key encoding both, the last previous hash key as well as the new transaction information.
\end{itemize}
\end{itemize}
Unfortunately, if it was known in advance which agent was going to confirm the transaction, then Alice could secretly collaborate with this agent and therefore manipulate the blockchain through an illegitimate extension.

To enforce a random choice of the confirming agent in the distributed network, the blockchain rules additionally require the newly generated hash key in b) to have a certain structure. For example, it may have to have a certain number of zeros at the beginning or end (or another similar specification). The distributed agents are then allowed to append the alpha-numeric string which is to be hashed by a personally chosen component called a \emph{nonce}. A confirmation is accepted as soon as one agent in the network manages to propose a nonce which results in a hash key of the required format.

 The competition in which all agents in the network simultaneously try to find a suitable nonce to complete the given task is referred to as \emph{mining}. The requirements of the new hash key can be chosen such that mining takes a certain average time (e.g., ten minutes) even for very fast computers. As long as a large number of participating agents in the distributed network have the computational ability to find the new hash key within the expected time, it is unlikely that one particular agent will be able to always dominate (and therefore manipulate) the process. 

Adding this detail to point iv. above, the more specifically phrased item b) becomes the following.

\begin{itemize}
\item[]
\begin{itemize}
\item[b)*] [...] and create a new hash key \emph{of specified format} encoding the combined string of \emph{three} items, the last previous hash key, the new transaction information, and a newly-mined nonce.
\end{itemize}
\end{itemize}

In the original bitcoin design, the completion of b)* is rewarded with a payment of (newly created) bitcoin, to ensure that participation in the confirmation process is desireable for a large number of distributed agents (and hence the name \emph{mining}).

Suppose that a random agent called Charlie is the first to successfully confirm Alice's proposed transaction of 30 bitcoins. Charlie then immediately broadcasts the new hash key together with Alice's new transaction to everyone in the distributed network -- a new \emph{block} is being added to the blockchain (i.e., the distributed ledger is updated). As soon as his synchronised version of the blockchain contains Alice's confirmed bitcoin transfer, Bob can hand over his bike to Alice and complete the sale of the underlying physical good. 

\section{Blockchain: Electronic Transaction Without Intermediary}

A certain technological complexity (e.g., as ensuring instant synchronisation of information across a large distributed network) aside, the blockchain design allows the creation of a secure distributed database for almost any type of information.

Three particular properties of the blockchain have to be kept in mind.

\subsection{Network Size}

It is crucial to have a large distributed network, which means agents have to be incentivised to participate in the mining process. In many cases, this may mean offering a small \emph{fee} for every successfully confirmed transaction (or newly created block).

\subsection{Blockchain Depth}

All agents participating in the distributed network will always update their version of the blockchain to the longest prevailing one. This means that if someone managed to manipulate (i.e., \emph{break}) the latest added block to re-route Alice's transfer after Bob has already seen it and then integrate it in a longer blockchain, then Bob might find himself handing over his bike even though his 30 bitcoins have actually been sent elsewhere.

The repeated hashing resulting from the addition of new blocks to the blockchain means that past transactions become more impenetrable as they sink deeper and deeper into the blockchain. If Bob, before handing over his bike, waits for a number of more blocks (the current standard is \emph{six}) to be added after his own transaction, than he can be certain that his receipt of money has been logged as a permanent and unchangeable part of the blockchain. 

\subsection{51$\%$ Attack}

While the name \emph{51$\%$ attack} is misleading (as the number 51 is of no particular relevance), this refers to the risk of having a maliciously intended agent or group of agents who dominate the aggregate computational power available in the distributed network -- they could, therefore, manipulate the addition of new blocks by consistently leading the confirmation process. In practice, the true impact of this is smaller than widely believed, as it primarily requires waiting for more blockchain depth before moving physical goods or making further money transfers -- creating a speed rather than a security problem.

\subsection{Theft}
The blockchain technology does not prevent theft of property. Especially for the anonymous design of the bitcoin, where an individual's money is stored as a collection of private keys in a bitcoin wallet, property can be stolen if the electronic wallet is accessed illegally and bitcoins are spent.

%\newpage 

{\bf DISCLAIMER}

The views expressed are those of the author and do not reflect the official policy or position of Record Currency Management.

\end{document}